\documentclass[reprint,twocolumn,showpacs,superscriptaddress,longbibliography,email,floatfix,aps,pra]{revtex4-2}

\usepackage{header}

\usepackage{tikz}
\definecolor{CB1}{HTML}{d7191c} %blue
\definecolor{CB2}{HTML}{fdae61}
\definecolor{CB3}{HTML}{ffffbf}
\definecolor{CB4}{HTML}{abd9e9}
\definecolor{CB5}{HTML}{2c7bb6} %red
\definecolor{CB6}{HTML}{32414b} %gray

\def \xgate{
    \tikz[every picture/.style={line width=1pt}, baseline=-3]{
        
        \draw [every picture] (0,0) -- (1,0);
        
        \filldraw [every picture][fill=CB5!60!white,draw=black] (0.25,-0.25) rectangle (0.75,0.25);
        \node at (0.5,0) {$\varphi_x$};
        \clip (0,-0.25) rectangle (1, 0.25);
        
    }
}

\def \nngate{
    \tikz[every picture/.style={line width=1pt}, baseline=6]{
    
        \def \spacing{0.6}
    
        \draw [every picture] (0,0) -- (1,0);
        \draw [every picture] (0,\spacing) -- (1,\spacing);
        \filldraw [every picture][fill=CB2!60!CB3,draw=black] (0.25,-0.125) rectangle (0.75,\spacing + 0.125);
        \node[rotate=-90] at (0.5, 0.5*\spacing) {$\varphi_{nn}$}; 
        
        \clip (0,-0.125) rectangle (1, \spacing);
        
    }
}

\def \xgatesmall{
    \tikz[every picture/.style={line width=1pt}, baseline=-3, transform canvas={scale=0.7}]{
        
        \draw [every picture] (0,0) -- (1,0);
        
        \filldraw [every picture][fill=CB5!60!white,draw=black] (0.25,-0.25) rectangle (0.75,0.25);
        \node at (0.5,0) {$\varphi_x$};

    }
}

\def \nngatesmall{
    \tikz[every picture/.style={line width=1pt}, baseline=1, transform canvas={scale=0.45}]{
    
        \def \spacing{0.6}
    
        \draw [every picture] (0,0) -- (1,0);
        \draw [every picture] (0,\spacing) -- (1,\spacing);
        \filldraw [every picture][fill=CB2!60!CB3,draw=black] (0.25,-0.125) rectangle (0.75,\spacing + 0.125);
        \node[rotate=-90] at (0.5, 0.5*\spacing) {$\varphi_{nn}$}; 
    }
}

\newcommand{\tubingen}{Institut f\"{u}r Theoretische Physik,  Universit\"{a}t T\"{u}bingen, Auf der Morgenstelle 14, 72076 T\"{u}bingen, Germany}
\newcommand{\mpipks}{Max Planck Institute for the Physics of Complex Systems, N\"othnitzer Str. 38, 01187 Dresden, Germany}

\begin{document}
\title{Machine learning of reduced quantum channels on NISQ devices}
\author{Giovanni Cemin}
\thanks{These two authors contributed equally to this work.}
\affiliation{\mpipks}
\author{Marcel Cech}
\thanks{These two authors contributed equally to this work.}
\affiliation{Institut f\"ur Theoretische Physik and Center for Integrated Quantum Science and Technology, Universit\"at T\"ubingen, Auf der Morgenstelle 14, 72076 T\"ubingen, Germany}
\author{Erik Weiss}
\affiliation{\tubingen}
\author{Stanislaw Soltan}
\affiliation{\tubingen}
\author{Daniel Braun}
\affiliation{\tubingen}
\author{Igor Lesanovsky}
\affiliation{Institut f\"ur Theoretische Physik and Center for Integrated Quantum Science and Technology, Universit\"at T\"ubingen, Auf der Morgenstelle 14, 72076 T\"ubingen, Germany}
\affiliation{School of Physics and Astronomy and Centre for the Mathematics and Theoretical Physics of Quantum Non-Equilibrium Systems, The University of Nottingham, Nottingham, NG7 2RD, United Kingdom}
\author{Federico Carollo}
\affiliation{Institut f\"ur Theoretische Physik and Center for Integrated Quantum Science and Technology, Universit\"at T\"ubingen, Auf der Morgenstelle 14, 72076 T\"ubingen, Germany}

\begin{abstract}
World-wide efforts aim at the realization of advanced quantum simulators and processors. However, despite the development of intricate hardware and pulse control systems, it may still not be generally known which effective quantum dynamics, or channels, are implemented on these devices. To systematically infer those, we propose  a neural-network algorithm approximating generic discrete-time  dynamics through the repeated action of an effective quantum channel. We test our approach considering time-periodic Lindblad dynamics as well as non-unitary subsystem dynamics in many-body unitary circuits. Moreover, we  exploit it to investigate cross-talk effects on the \texttt{ibmq\_ehningen} quantum processor, which showcases our method as a practically applicable tool for inferring quantum channels when the exact nature of the underlying dynamics on the physical device is not known \textit{a priori}. While the present approach is tailored for learning Markovian dynamics, we discuss how it can be adapted to also capture generic non-Markovian discrete-time  evolutions. 
\end{abstract}

\maketitle

% -------------------------------------------------------------------------------------------------------------------------------------------- %

\section{Introduction} 

As large-scale fault-tolerant quantum machines may still be years away  \cite{Monroe2014,Linke2017,Chao2017,Cerezo2020,Cohen2021}, quantum simulation and computation tasks can nowadays only be performed on noisy intermediate-scale quantum (NISQ) devices \cite{barratt2021,daley2022}. The latter are unavoidably affected by noise and errors, which ultimately limit their applicability  \cite{Preskill2018,Bharti2022,Sun2021,Brandhofer2022,Cech2023,Dalton2024}.
Given their intricate structure and their coupling with the surroundings, it is not often fully clear which quantum dynamics, be it unitary or non-unitary, is actually implemented on these machines. 
A complete characterization accounting for all the microscopic degrees of freedom is prohibitively complex. An effective description, which solely considers ``logical" degrees of freedom, i.e., those relevant for physically interesting few-qubit observables, may thus prove valuable in shedding light on the nature of the effective dynamical behavior of quantum devices. For example, a time-independent Lindblad master equation, encoding a simple instance of open quantum dynamics \cite{Lindblad1976,Gorini1976,Breuer2002,Manzano2020}, can provide insights into different error mechanisms  \cite{Samach2022,Berg2023,DiBartolomeo2023,StilckFranca2024}. An additional difficulty stems from the fact that digital quantum processors exploit fast time-dependent pulses to optimize the speed and the fidelity of their logical gates \cite{McKay2018,Cross2022,Machnes2018,DiBartolomeo2023}. Therefore, it is unclear whether an effective (time-dependent) Lindblad master equation even exists  \cite{Wolf2008,Wolf2008a,Schnell2020,Chruscinski2022,Nestmann2021,Gaikwad2024}. As such, characterizing the dynamics in NISQ devices is still, to a large extent, an open problem. 

Various methods have been put forward to address this challenge \cite{Eisert2020,Harper2023,Ilin2024,Duvenhage2021}, including, for example, gate set tomography \cite{Greenbaum2015,BlumeKohout2017,Nielsen2021,Madzik2022,BlumeKohout2022,Brieger2023}, Lindblad tomography \cite{Samach2022,Berg2023,Varona2024}, non-Markovian processes tomography \cite{White2022,Agarwal2024}, Hamiltonian learning \cite{Wang2015,Wilde2022,Haah2024,Hangleiter2024,StilckFranca2024}, and very recently a pattern-based approach \cite{Weiss2024}.
Machine learning (ML) approaches appear to be particularly suited for this task \cite{Han2021,Mohseni2022,Braccia2022,Huang2023,Gebhart2023}. However, proposed methods either rely on an \textit{ad-hoc} ansatz, require data which are not experimentally accessible, or lack physical interpretability (which is actually becoming highly desirable \cite{Link2023}).

\begin{figure}[t]
    \centering
    \includegraphics{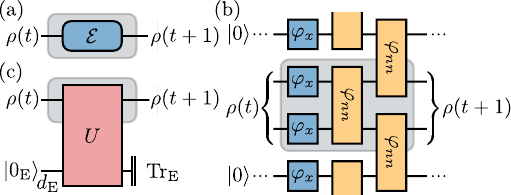}
    \caption{\textbf{Learning quantum channels.} (a) A general quantum system evolving in discrete time is described by an update rule for its density matrix, $\rho(t) \to \rho(t+1)$. Here, we infer an {\it optimal} quantum channel $\mathcal{E}$, approximating the evolution as $\rho(t+1)=\mathcal{E}[\rho(t)]$.
    (b) Machine learning the effective discrete-time subsystem dynamics (gray box) in a unitary circuit. In the shown example, the circuit consists of single-qubit rotations around the $x$-axis (rotation angle $\varphi_x$) and controlled phase gates on adjacent qubits (phase $\varphi_{nn}$). (c) The quantum channel (gray box) is learned by exploiting a  \textit{collision model}, which describes the system dynamics as emerging from a unitary interaction, $U$, with an environmental particle having Hilbert space dimension $d_\mathrm{E}$.
    }
    \label{fig:fig1}
\end{figure}

In this work, we propose a simple method capable of approximating a generic discrete-time dynamics [cf.~Fig.~\ref{fig:fig1}(a,b)] through the repeated application of an effective quantum channel \cite{Nielsen2010}. Here, the channel is constructed by considering a unitary interaction between the system of interest and fictitious auxiliary degrees of freedom \cite{Stinespring1955,Ciccarello2022,Cech2023}, as sketched in Fig.~\ref{fig:fig1}(c). We then employ machine-learning techniques to infer the appropriate structure of this interaction. Such a representation of a quantum channel allows us to encode physical constraints directly into the neural-network ansatz and significantly facilitate the learning process \cite{Kraus1983,Nielsen2010,Ahmed2023,Choi1975,Audenaert2002,Kofman2009,Baldwin2014,Banchi2020,Teo2020,Acharya2023,Ilin2024}. 
To illustrate its effectiveness, we benchmark our approach with two different settings: a periodic open-system dynamics generated by a time-dependent Lindblad generator and the reduced dynamics of a subsystem embedded in a unitary many-body circuit, shown in Fig.~\ref{fig:fig1}(a,b) respectively. The first scenario allows us to unambiguously demonstrate the power of our method while the second one allows us to explore its capability to infer  non-Markovian  circuit dynamics. Importantly, the  focus is not on learning the full dynamics of a quantum system, which may become impractical in many-body settings, but rather on characterizing in an interpretable way the dynamics of reduced subsystems embedded in a larger one.  

Finally, we show an application to the \texttt{ibmq\_ehningen} quantum processor.  In contrast to idealized frameworks, this entails dealing with practical aspects, including imperfect knowledge of the quantum state due to limited number of measurements, i.e., shot noise.
Using data generated by the NISQ device, we can nevertheless efficiently infer a quantum channel, which accurately captures qubit cross-talk in the inspected circuit. 
These findings demonstrate the usefulness of our approach for characterizing quantum dynamics on existing devices. In these settings, the learned channel may find applications in the development of error mitigation strategies \cite{Sidi2003,Temme2017,Krebsbach2022,RodriguezVega2022,Berg2023,Kim2023}.

%\section{Quantum channels and machine learning}
\section{Methods}
\subsection{Quantum channels}

Let us consider a generic quantum system, with finite Hilbert-space dimension $d$, evolving in discrete time. The state of the system is described in terms of a density matrix $\rho$, providing the expectation value of any quantum observable $O$, as $\langle O\rangle ={\rm Tr}\left(\rho \,  O\right)$. Assuming the system to be subject to a Markovian discrete-time dynamics \cite{Wolf2008,Breuer2009,Benatti2013,Roos2020,Manzano2020}, the density matrix evolves according to the iterative equation $\rho(t + 1) = \mathcal{E}[\rho(t)]$, with solution $\rho(t) = \mathcal{E}^t[\rho(0)]$ \cite{Nielsen2010}. The quantum channel $\mathcal{E}$, encoding, e.g., a genuine dynamics or a generic quantum operation, must have the form \cite{Nielsen2010}
\begin{align}
    \mathcal{E}[\rho] = \sum_{k = 0}^{d_\mathrm{E}-1} K_k \rho K_k^\dagger \, ,
    \label{eq:Kraus_map}
\end{align}
with the $d_{\rm E}$ operators $\{ K_k \}_{k=0}^{d_\mathrm{E}-1}$ being the so-called Kraus operators. The structure in Eq.~\eqref{eq:Kraus_map} ensures that  $\mathcal{E}$ is  completely positive, while the additional condition $\sum_{k = 0}^{d_\mathrm{E}-1} K_k^\dagger K_k  = \mathds{1}$ guarantees preservation of the trace, i.e., ${\rm Tr}\left(\mathcal{E}[\rho]\right)={\rm Tr}\left(\rho\right)$. Both properties are required for $\mathcal{E}$ to represent a physical quantum evolution \cite{Nielsen2010,Wood2015,Manzano2020,Chruscinski2022,Ahmed2023}. Note that, for $d_\mathrm{E} = 1$, Eq.~\eqref{eq:Kraus_map}  can only  describe unitary operations, while any possible physical (open-system) dynamics  can be represented by considering, at most, $d_{\rm E}=d^2$ Kraus operators  \cite{Nielsen2010,Manzano2020}. The minimal $d_{\rm E}$ required to represent a given quantum channel is the so-called Kraus rank \cite{Nielsen2010}.

The Stinespring dilation theorem \cite{Stinespring1955} states that a completely positive and trace-preserving quantum channel can be written as 
\begin{align}
    \mathcal{E}[\rho] = \Tr_\mathrm{E} \left[ U ( \rho \otimes \rho_\mathrm{E} ) U^\dagger \right] \, . 
    \label{eq:Stinespring_ansatz}
\end{align}
This relation provides a useful  interpretation of $\mathcal{E}$ in terms of a collision model: the system dynamics emerges from the unitary interaction of the latter with an auxiliary {\it environmental} particle, having Hilbert space dimension $d_{\rm E}$ \cite{Stinespring1955,Nielsen2010,Ciccarello2022,Cech2023}. The auxiliary particle can be initialized in a reference pure state, $\rho_\mathrm{E} = \ketbra{0_{\rm E}}$, and the unitary operator $U$ acts on both the system and the auxiliary particle. The trace ${\rm Tr_E}$ over the environmental degrees of freedom, represented, e.g., through the basis $\{{\ket{k_{\rm E}}}\}_{k=0}^{d_{\rm E}-1}$, provides the reduced state of the system after the collision, and   generates the Kraus operators  $K_k=\bra{k_{\rm E}}U\ket{0_{\rm E}}$ appearing in Eq.~\eqref{eq:Kraus_map}. 

\subsection{Machine learning algorithm}
The question that we address in the following is whether, given a generic discrete-time dynamics, it is possible to infer the form of an optimal quantum channel $\mathcal{E}$ approximating it.
To this end, we exploit the Stinespring dilation theorem \cite{Stinespring1955} and choose the ansatz outlined in Eq.~\eqref{eq:Stinespring_ansatz} [see also Fig.~\ref{fig:fig1}(c)]. We parametrize $U$ through a neural-network and constrain the output to be a unitary matrix by means of Pytorch's \texttt{torch.nn.utils.parametrizations.orthogonal} \cite{Paszke2019}. This last step is necessary to ensure that our model represents a physical, i.e., completely positive and trace preserving, quantum channel. 
The matrix $U$ has dimensions $(d \, d_E) \times (d \, d_E)$, where $d$ and $d_E$ are the dimension of the system Hilbert space and the environmental Hilbert space, respectively.  

Throughout this work, we focus on systems composed of $L$ qubits and the corresponding $d=2^L$ dimensional Hilbert space. 
Here, the system state $\rho$ can be characterized in terms of the expectation values of Pauli strings $\{F_j\}_{j=1}^{d^2} = \{\mathds{1} \otimes ... \otimes \mathds{1}, ..., \sigma_z \otimes ... \otimes \sigma_z \}$ \cite{Mazza2021,Carnazza2022,Cemin2024}, as $\rho = (\mathds{1} + \sum_{j=2}^{d^2} F_j v^j)/d$, where $v^j = \expval{F_j} = \Tr[F_j \rho]$. The elements $v^j$ are the entries of the so-called coherence vector $\mathbf{v} = (v^j)_{j = {1, ..., d^2}}$. 
Our model takes as input the state at time $t$ (represented  via  the coherence vector) and calculates the state at time $t+1$ by providing the updated coherence vector via the relations 
\begin{align*}
    {v}^i_{\rm ML} (t+1) = \Tr\Big[F_i \mathcal{E}[\rho_{\rm ML}(t)]\Big] \,,
\end{align*}
with 
\begin{align*}
    \rho_{\rm ML}(t) = \frac{1}{d} \Big[\mathds{1} + \sum_{j=2}^{d^2} F_j v^j_{\rm ML}(t)\Big]\, .
\end{align*}
For convenience, we consider the initial states to be pure product states, where the single-qubit states are sampled from the Haar measure \cite{Meckes2019,Qiskit2019}. 
The coherence vector ${\bf v}_{\rm ML}(t)$ is thus obtained by propagating this initial condition using the neural-network model up to the corresponding time. To train the model, we first gather the training data set: we estimate (either from numerical simulations or from measurement outcomes) the coherence vectors ${\bf v}(t)$, associated with the exact quantum dynamics, for all the considered times and for different initial conditions. 
We then minimize the loss function 
$$
\mathbb{L} 
= \mathds{E}\big\{|\mathbf{v}_\mathrm{ML}(t) - \mathbf{v}(t)|^2\big\}=d\, \mathds{E}\big\{{\rm Tr}\left[(\rho_\mathrm{ML}(t) - \rho(t))^2\right]\big\}\, ,
$$
that is proportional to the average square-distance between the exact density matrix $\rho(t)$ and the one predicted by a neural network, $\rho_\mathrm{ML}(t)$. The symbol $\mathds{E}$ denotes the average over $M$ dynamical trajectories $\{\rho(t)\}$, generated starting from a Haar random distributed initial state, and sampling at discrete times $t$. 
To perform the minimization we utilize the Adam optimizer \cite{Kingma2017}.

Note that, in addition to standard hyper-parameters in the training, the environmental dimension $d_{\rm E}$ can also be considered as a hyper-parameter of the model. In the following, we explore values of $d_{\rm E}$ different from  the Kraus rank required by the exact dynamics, to explore how this impacts the learning process \cite{Gross2010,Flammia2012,Rodionov2014}. 

After the training, we validate the learned channel against data generated from $r=10$ additional initial conditions, and calculating the error
\begin{align}
	\epsilon = \frac{1}{r} \sum_{j=1}^r \frac{1}{T} \sum_{t = 1}^T \frac{{\rm Tr}\left[( \rho_\mathrm{ML}^j(t) - \rho^j(t))^2\right]}{{\rm Tr}\left[(\rho^j(t))^2 \right]} \, .
    \label{eq:error_function}
\end{align}
Here, the superscript $j$ denotes trajectories from different initial conditions while $T$ is the considered final time.

\begin{table}[b]
\centering
\caption{Hyper-parameters used in the training processes of the different examples. \texttt{lr} stands for the learning rate of the Adam optimizer, \texttt{gamma} is the multiplicative factor for the learning rate decay,  \texttt{batch\_size} the size of the batch utilized in each epoch, and \texttt{n\_epochs} is the total number of epochs.}
\begin{tabular}{m{2.25cm} || >{\centering\arraybackslash}p{1cm} | >{\centering\arraybackslash}p{1cm} | >{\centering\arraybackslash}p{1.8cm} | >{\centering\arraybackslash}p{1.5cm}}
    Model for & \texttt{lr} & \texttt{gamma}& \texttt{batch\_size} & \texttt{n\_epochs} \\ 
    \hline
    Figs.~\ref{fig:benchmarking_stroboscopic_dynamics},\ref{fig:further_insights_benchmarking_stroboscopic_dynamics} & $0.002$ & $0.999$ & $128$ & $400$ \\  
    Figs.~\ref{fig:subsystem_dynamics},\ref{fig:fig_s4} & $0.002$ & $0.999$ & $128$ & $400$ \\ 
    Figs.~\ref{fig:ibmq_ehningen},\ref{fig:fig_s5},\ref{fig:fig_s6}: &&&\\
    $\quad$Pre-training & $0.001$ & $1.000$ & $256$ & $300$\\
    $\quad$Training & $0.001$ & $0.980$ & $256$ & $300$ \\
    Fig.~\ref{fig:fig_s2} & $0.002$ & $0.999$ & $128$ & $400$ \\ 
    Fig.~\ref{fig:fig_s3} & $0.002$ & $0.999$ & $128$ & $400$ 
\end{tabular}
\label{table:1}
\end{table}

In Table~\ref{table:1}, we also provide the main hyper-parameters used during the training process. 

\subsection{Addressing non-Markovian dynamics}
Before we proceed, we comment on our underlying assumption of Markovianity [see also sketch in Fig.~\ref{fig:fig1}(a)]. On the one hand, while we can train the network using any dynamics, this assumption clearly represents a limitation on its ability to capture generic evolutions. In return, however, compared to less constrained representations \cite{Kraus1983,Nielsen2010,Ahmed2023,Choi1975,Audenaert2002,Kofman2009,Baldwin2014,Banchi2020,Teo2020,Acharya2023,Duvenhage2021}, the structure in Eq.~\eqref{eq:Stinespring_ansatz} offers enhanced interpretability of the learned channel, whose effects are encoded in the unitary interaction $U$. Moreover, our method can be adapted to include the learning of non-Markovian discrete-time dynamics. Instead of learning a single channel $\mathcal{E}$ implementing single-time updates, one could learn the different quantum channels mapping the system from the initial time to each later time $t$. The knowledge of this family of channels, which must all have the form given in Eqs.~\eqref{eq:Kraus_map}-\eqref{eq:Stinespring_ansatz} due to physical constraints, permits the analysis of whether the dynamics is time-dependent and whether it is divisible in terms of completely positive channels  \cite{Wolf2008a,Breuer2009,Benatti2013,Roos2020,Samach2022}.

% -------------------------------------------------------------------------------------------------------------------------------------------- %
\section{\label{sec:Applications}Applications}
\subsection{Time-periodic Lindblad  dynamics} 

As a first benchmark, we  consider a single qubit evolving through the master equation $\dot{\rho}(s) = \mathcal{L}(s) \left[\rho(s)\right]$, governed by the time-periodic Lindblad generator \cite{Schnell2020}
\begin{align}
	 \mathcal{L}(s)[\rho] = -i[H(s), \rho] + \gamma \left(J \rho J^\dagger  - \frac{1}{2} \{ J^\dagger J, \rho \} \right) \, .
	\label{eq:time_periodic_Lindblad}
\end{align}
The system Hamiltonian is $H(s) = E_z \sigma_z + E_x \cos(\omega s) \sigma_x$, and we consider $E_x = E_z / 2$, $\omega = E_z$, $\gamma = 0.01\,E_z$ and $J = \ketbra{0}{1}$. Since we want to learn an effective discrete-time dynamics, we focus on the stroboscopic evolution generated by the one-period map $\mathcal{E} = \mathcal{T} e^{ \int_0^{2\pi / \omega} {\rm d}s \, \mathcal{L}(s)}$, with $\mathcal{T}$ being the time-ordering operator \cite{Breuer2002}. 
A general question in these settings is whether the stroboscopic dynamics can be obtained by means of an effective time-independent Floquet Lindblad generator $\mathcal{L}_\mathrm{F}$, as $\mathcal{E} =e^{{\frac{2\pi}{\omega}\mathcal{L}_\mathrm{F}}}$ \cite{Andersson2007,Wolf2008,Wolf2008a,Roos2020,Schnell2020,Chruscinski2022,Baecker2024}. To explore this aspect and its impact on the learning of the effective dynamics, we also employ the Lindblad approximator of Ref.~\cite{Cemin2024} to infer a Floquet generator $\mathcal{L}_\mathrm{F}$ from the stroboscopic evolution.

\begin{figure}[t]
    \centering
    \includegraphics{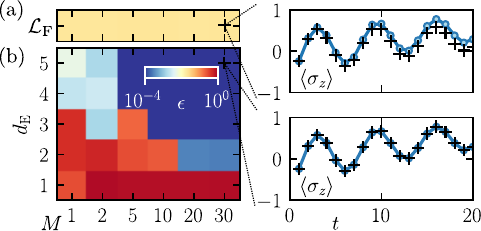}
    \caption{\textbf{Time-periodic Lindblad dynamics.} (a) Error measure, as defined in Eq.~\eqref{eq:error_function},  associated with the inferred  Lindblad approximation, $\mathcal{L}_\mathrm{F}$, obtained with the method of Ref.~\cite{Cemin2024}. The right panel shows the inferred (black crosses) and the exact (blue circles) dynamics  for a representative expectation value.  
    (b) Error measure [cf.~Eq.~\eqref{eq:error_function}], associated with the learned quantum channel $\mathcal{E}$, as a function of the number $M$ of considered initial conditions [as in panel (a)] and of the dimension $d_{\rm E}$ of the environmental  particle.  The right panel shows the inferred (black crosses) and the exact (blue circles) dynamics  for a representative expectation value.  The stroboscopic time is given  in units of $2\pi/\omega$ (see main text).
    }
    \label{fig:benchmarking_stroboscopic_dynamics}
\end{figure}

In Fig.~\ref{fig:benchmarking_stroboscopic_dynamics}(a), we observe, both from the error measure $\epsilon$ [cf.~Eq.~\eqref{eq:error_function}] and from the time-evolution of $\expval{\sigma_z}$, a discrepancy between the exact dynamics and the one associated with the learned Floquet generator $\mathcal{L}_{{\rm F}}$.
This occurs even though the considered parameters are such that a well-defined $\mathcal{L}_{{\rm F}}$ exists (see following discussion).
Fig.~\ref{fig:benchmarking_stroboscopic_dynamics}(b) displays results on the learned quantum channel $\mathcal{E}$. As shown, for sufficiently large training data sets and auxiliary dimension $d_\mathrm{E} \geq 3$, the dynamics is accurately inferred. Here, we note that $d_\mathrm{E}$ can stabilize the training process even for small training data sets (see $d_\mathrm{E} = 4, 5$, $M = 5$). 

\begin{figure}
    \centering
    \includegraphics{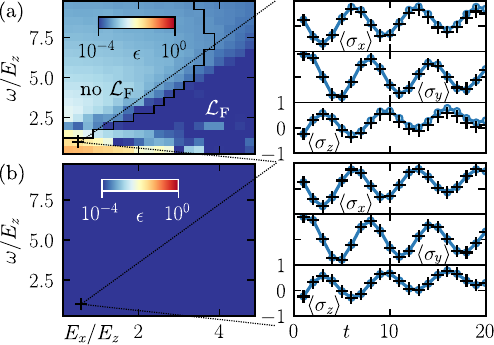}
    \caption{\textbf{Floquet generators and quantum channels of time-periodic Lindblad dynamics.} (a)~Error measure, as defined in Eq.~\eqref{eq:error_function}, related to the inferred effective Lindblad dynamics, obtained using the method introduced in Ref.~\cite{Cemin2024}. For each combination of $\omega / E_z$ and $E_x / E_z$ we train on $M = 100$ different trajectories and then compute the error $\epsilon$ using $r=10$ out-of-sample trajectories. The solid black line separates the regime where an effective time-independent Lindblad description of the underlying time-dependent Lindblad in Eq.~\eqref{eq:time_periodic_Lindblad} exists from the one in which there is no $\mathcal{L}_{\rm F}$ \cite{Wolf2008,Schnell2020}.
    For the parameters $(E_x/ E_z, \omega / E_z) = (0.5, 1)$ used in Fig.~\ref{fig:benchmarking_stroboscopic_dynamics}, we show the time-evolution (blue circles - exact data, black crosses - learned dynamics) from an initial state in the validation  set. (b)~Error measure associated with the learned channels with $d_\mathrm{E} = 4$ that were trained on $M=30$ different trajectories. As apparent, the learned channel better captures the underlying dynamics in all regimes, including the region where no time-independent effective Lindblad description is present.
    }
    \label{fig:further_insights_benchmarking_stroboscopic_dynamics}
\end{figure}

The results presented in Fig.~\ref{fig:benchmarking_stroboscopic_dynamics}(a) seem to suggest that no well-defined Floquet generator $\mathcal{L}_{{\rm F}}$ exists.
To better understand this, we now carry out a more thorough exploration of the time-dependent Lindblad generator $\mathcal{L}(s)$ in Eq.~\eqref{eq:time_periodic_Lindblad} for different combinations of $\omega / E_z$ and $E_x / E_z$. In Fig.~\ref{fig:further_insights_benchmarking_stroboscopic_dynamics}(a) we show with a black solid line the separation between the region where a Floquet generator exists ($\mathcal{L}_{\rm F}$) and a region where it does not exist (no $\mathcal{L}_{\rm F}$). 
To determine this boundary, we follow the approach outlined in Refs.~\cite{Wolf2008,Schnell2020}. In our case, the existence of the Floquet generator $\mathcal{L}_\mathrm{F}$ is conditioned on a Hermitian logarithm of the one-period map, i.e., $\mathcal{L}_\mathrm{F} = \frac{\omega}{2\pi} \log \mathcal{E}$, being conditionally completely positive \cite{Evans1977}. 
In the same plot, we show the error obtained from extracting a Lindblad approximation using the method presented in Ref.~\cite{Cemin2024}. The comparison highlights that the parameters $(E_x/E_z, \omega / E_z) = (0.5, 1)$ considered in Fig.~\ref{fig:benchmarking_stroboscopic_dynamics} in principle give rise to dynamics explained by a Floquet generator $\mathcal{L}_\mathrm{F}$, which however cannot be discovered using the Lindblad approximator of Ref.~\cite{Cemin2024}. This failure may be related to the parameters of the networks being stuck in a local minimum of the loss function [see the yellow region in Fig.~\ref{fig:further_insights_benchmarking_stroboscopic_dynamics}(a) and the discrepancy between the blue circles and the black crosses in the time-evolution on the right  of panel (a)], which could be associated with the non-convex geometry of the set of channels generated by a Lindblad master equation \cite{Wolf2008}. Conversely, using the above proposed algorithm to directly infer the quantum channel, we achieve a very high accuracy throughout the whole parameter regimes [see blue region indicating $\epsilon < 10^{-4}$ in Fig.~\ref{fig:further_insights_benchmarking_stroboscopic_dynamics}(b)]. This clearly highlights a benefit of our method that is able to infer an accurate quantum channel across all the parameters we explored.

\subsection{Reduced subsystem dynamics} 

\begin{figure}  
    \centering
    \includegraphics{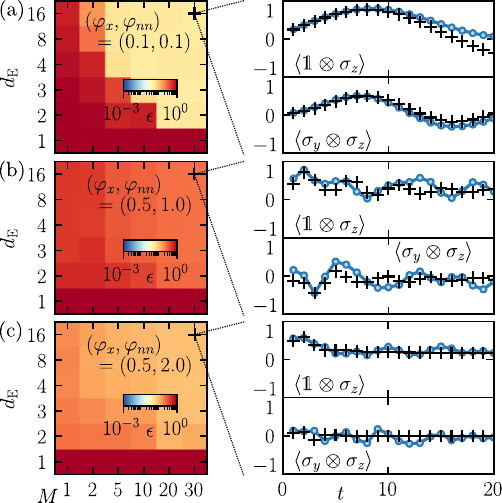}
    \caption{\textbf{Reduced subsystem dynamics.} Reduced dynamics of a two-qubit subsystem embedded in a circuit of $L=14$ qubits, as shown in Fig.~\ref{fig:fig1}(b). 
    (a) Error measure, as defined in Eq.~\eqref{eq:error_function}, associated with the learned quantum channel $\mathcal{E}$ approximating the subsystem dynamics. The error is given as a function of the number $M$ of considered initial conditions and of the dimension $d_{\rm E}$ of the environmental particle. The circuit parameters are $(\varphi_x, \varphi_{nn}) = (0.1, 0.1)$. The right plots provide the inferred (black crosses) and the exact dynamics (blue circles) for representative expectation values.
    (b,c) Same as in panel (a) for circuit parameters $(\varphi_x, \varphi_{nn})=(0.5, 1)$ and $(\varphi_x, \varphi_{nn})=(0.5, 2)$ respectively.}
    \label{fig:subsystem_dynamics}
\end{figure}

The previous example establishes that our approach can accurately infer quantum channels within a controlled  Markovian setting. Next, we investigate whether it can provide an accurate approximation in the case of non-Markovian dynamics \cite{Roos2020,Baecker2024}.  
We focus on the time evolution of a subsystem embedded in a many-body system. The dynamics of the full system is described by a unitary quantum circuit, depicted in Fig.~\ref{fig:fig1}(b), built from the unitary gates
\begin{align}
    \xgate = e^{-i \sigma_x \varphi_x}\, , \hspace{1cm} \nngate = e^{i  ({n} \otimes  {n}) \varphi_{nn}} \, ,
    \label{eq:gates_subsystem_dynamics}
\end{align}
with ${n} = (\mathds{1} - \sigma_z) / 2=\ketbra{1}{1}$. A single time-update consists in the application of the rotation \hspace{0.05cm}$\xgatesmall$\hspace{0.7cm} around the $x$-axis (angle $\varphi_x$) on each qubit followed by  the application of the controlled phase gate \hspace{0.05cm}$\nngatesmall$\hspace{0.5cm} (phase $\varphi_{nn}$) on each pair of  neighboring qubits [cf.~Fig.~\ref{fig:fig1}(b)] \cite{Krantz2019,Collodo2020,Xu2020,Alexander2020}. 

The task is to learn the reduced dynamics of two adjacent qubits embedded in a quantum circuit of $L$ qubits (with periodic boundary conditions)  [cf.~Fig.~\ref{fig:fig1}(b)]. The two qubits are initialized in a random state while the remainder of the circuit in state $\ket{0}$.
For generic values of the parameters $\varphi_x$ and $\varphi_{nn}$, such dynamics is not expected to be Markovian. Nonetheless, an optimal channel  $\mathcal{E}$, which approximates the discrete-time evolution, can be obtained. We train the network for times $t =1,... 10$ and validate the learned dynamics up to $T = 20$. Validating over a longer time window allows us to better probe whether the dynamics is well approximated by a time-independent channel $\mathcal{E}$. 

The results are reported in Fig.~\ref{fig:subsystem_dynamics}. We consider three parameter regimes: $(\varphi_x, \varphi_{nn}) = (0.1, 0.1)$, $(\varphi_x, \varphi_{nn})=(0.5, 1)$ and $(\varphi_x, \varphi_{nn})=(0.5, 2)$. A direct comparison shows that the dynamics in the second case [cf.~Fig.~\ref{fig:subsystem_dynamics}(b)] is not accurately captured by the Markovian approximation, even at short times. Conversely, for the first parameter pair, a discrepancy between the learned and the exact dynamics primarily arises at later times [cf.~Fig.~\ref{fig:subsystem_dynamics}(a)]. This suggests that for   $(\varphi_x, \varphi_{nn})=(0.5, 1)$ non-Markovian effects are more pronounced than for $(\varphi_x, \varphi_{nn}) = (0.1, 0.1)$. This may seem obvious, at first, due to a larger coupling $\varphi_{nn}$ of the system with the rest of the circuit. 
However, we observed that for even larger values of $\varphi_{nn}$ [cf.~Fig.~\ref{fig:subsystem_dynamics}(c)], the approximation through the inferred Markovian model can improve again, which may be due to faster mixing effects in the unitary circuit dynamics. 

From a more general point of view, the different qualitative and quantitative agreement of the Markovian approximations also sheds light on dynamics beyond the subsystem alone. 
While small interactions [see Fig.~\ref{fig:subsystem_dynamics}(a)] require a longer period of time for correlations to emerge, larger interactions [see Fig.~\ref{fig:subsystem_dynamics}(b)] promote notable back-action between the subsystem and its ``reservoir'' (here, the remaining $12$ qubits). However, when interactions increase further [see Fig.~\ref{fig:subsystem_dynamics}(c)], the whole system may behave approximately uniformly such that a Markovian description incorporating only the subsystem applies again.

\subsection{Dynamics on the IBM quantum processor} 

\begin{figure}
    \centering
    \includegraphics{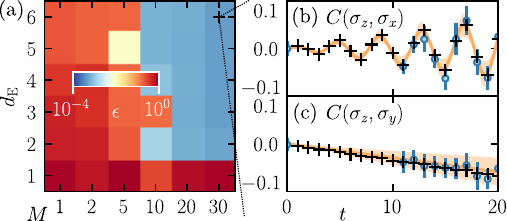}
    \caption{\textbf{Parallel $\sqrt{X}$ dynamics on  \texttt{ibmq\_ehningen}}.  (a) Error measure, as defined in Eq.~\eqref{eq:error_function}, associated with the learned quantum channel $\mathcal{E}$, for different numbers of initial conditions $M$ and dimensions of the environmental particle $d_{\rm E}$. 
    (b,c) We show the learned (black crosses) and the exact (blue circles) evolution  of the Pearson correlation coefficient [cf.~Eq.~\eqref{eq:Pearson_Correlation_Coefficient}] for two different pairs of observables. The exact dynamics is derived by running the circuit (see main text) on the quantum computer and estimating the density matrix from a finite measurement statistics.  The solid (yellow) line and the shaded region provide the Pearson correlation coefficient for the model in Eq.~\eqref{eq:ZZ_coupling_Hamiltonian}, with $V \in (-0.003, -0.001)$.
     }
    \label{fig:ibmq_ehningen}
\end{figure}

In the following, we explore quantum channels that are implemented on the 27-qubit \texttt{ibmq\_ehningen} NISQ processor. 
In this setting, the system state is not perfectly known  but can only be estimated from a finite number of measurements. Going beyond synthetic data, which can be generated on a classical computer with arbitrary precision, thus demonstrates that our method is capable to cope with shot noise, inherent to any real device  \cite{McKay2018,Cross2022}.

We consider a simple circuit, consisting of the repeated application of the native $\sqrt{X}$ gate ---corresponding to the case in Fig.~\ref{fig:fig1}(b) with $\varphi_x = \pi / 4$ and $\varphi_{nn} = 0$--- on two neighboring qubits (namely qubit $4$ and qubit $7$), while leaving the remaining ones idle. 
This dynamics should map product states into product states \cite{Nielsen2010,DiBartolomeo2023}. However, it has been observed that cross-talk effects between neighboring qubits lead to the emergence of a spurious $ZZ$-coupling \cite{Murali2020,Sung2021,Heunisch2023,Ketterer2023,Perrin2024}, responsible for the generation of correlations between the otherwise uncoupled qubits. 
Even though the ideal circuit only consists of $\sqrt{X}$ gates, the one occurring on the quantum computer is thus approximately described by the gate
\begin{align}
	G= \exp{-i \left[ \frac{\pi}{4} (\sigma_x \otimes \mathds{1} + \mathds{1} \otimes \sigma_x) + V \sigma_z \otimes \sigma_z \right]} \, ,
	\label{eq:ZZ_coupling_Hamiltonian}
\end{align}
with $V$ parametrizing the spurious coupling  \cite{Gambetta2012,Samach2022,Zhao2022}.  
In order to quantify such a cross-talk effect, we compute the Pearson correlation coefficient between observables pertaining to the different qubits, $\sigma_{\alpha_1}$ and $\sigma_{\alpha_2}$,  respectively.   The  coefficient is defined as  \cite{Giorgi2019,Carroll2022,Harper2023} 
\begin{align}
    C(\sigma_{\alpha_1}, \sigma_{\alpha_2}) = \frac{\expval{\sigma_{\alpha_1} \otimes \sigma_{\alpha_2}} - \expval{\sigma_{\alpha_1}} \expval{\sigma_{\alpha_2}}}{\sqrt{\expval{\sigma_{\alpha_1}^2} - \expval{\sigma_{\alpha_1}}^2} \sqrt{\expval{\sigma_{\alpha_2}^2} - \expval{\sigma_{\alpha_2}}^2}} \, .
    \label{eq:Pearson_Correlation_Coefficient}
\end{align}

The data acquisition from the \texttt{ibmq\_ehningen} quantum processor was done by generating $40$ different initial states for the two qubits of interest, by applying the native $\sqrt{X}$ gates on them 20 times and by performing measurements for $t=11,12\dots 20$. We run a total of $20\,000$ shots in each of the $9$ necessary Pauli bases. The number of measurement bases could be reduced considering  mutually unbiased observables \cite{Lawrence2002,Romero2005}. However, this would require the application of additional two-qubit gates, introducing further errors in the readout phase
 \cite{Smith2019,Leontica2021}. 
We separate the obtained data into $M=30$ initial conditions dedicated to the training, and $r=10$ initial conditions dedicated to validation.
To construct the training data set, instead of averaging over all obtained measurements to get a single estimate of the density matrix, we partition the entire set into $10$ subsets and then average within the subsets. As a result, the individual data points are subject to a larger statistical error but the size of the training set is effectively ``increased". This procedure proves valuable since the algorithm can tolerate high shot noise \cite{Cemin2024} and since a larger data set is beneficial for the training.
To further facilitate the process, we pre-train the algorithm by estimating the state at $t=10$ and learning in the time window $t=11,12,\dots 20$.  In this way, the learning is more stable due to the fewer applications of $\mathcal{E}$ needed to propagate to training times.

The validation error, shown in  Fig.~\ref{fig:ibmq_ehningen}(a), demonstrates that it is possible to learn a quantum channel that accurately reproduces the circuit dynamics on the \texttt{ibmq\_ehningen}. This is possible even with relatively small data sets, which is important for applications such as calibration and error mitigation \cite{Sidi2003,Temme2017,Krebsbach2022,RodriguezVega2022,Berg2023,Kim2023}. In particular, the inferred channel captures the weak, yet evident, cross-talk effects as shown by the nonzero Pearson correlation coefficient in Fig.~\ref{fig:ibmq_ehningen}(b,c). Despite the emergence of the spurious coupling, the $\sqrt{X}$ gate on the \texttt{ibmq\_ehningen} is close to the ideal one (see Appendix~\ref{sec:ibmq_ehningen}). The large purity of the quantum state observed from the data of the NISQ device is  reflected in the ability of the network to learn the dynamics with relatively small $d_\mathrm{E}$. 

% -------------------------------------------------------------------------------------------------------------------------------------------- %
\section{Conclusions and outlook} 
We have presented a machine-learning algorithm that is capable of approximating arbitrary discrete-time dynamics. The latter is based on a collision-model representation of  quantum channels, which guarantees  its physical consistency. We have benchmarked the method using synthetic data obtained from numerical simulations, and we have further shown its usefulness in understanding effective dynamics on the   \texttt{ibmq\_ehningen} quantum processor. 
This analysis further demonstrates its capability to work with real devices, for which only a finite number of measurements is available.
In this regard, it would be interesting to explore how our neural-network method performs when training solely from partial information of the density matrix, i.e., from measurements on a single or on an incomplete set of bases. 
Regarding future works, we also expect our framework to be valuable in investigating the emergence of non-Markovian dynamics on NISQ devices.

The code and the data that support the findings of this work are available on Zenodo \cite{ZenodoData}.\\

% -------------------------------------------------------------------------------------------------------------------------------------------- %

\acknowledgments
We thank Francesco Carnazza, Mari Carmen Bañuls, Michele Coppola and Zala Lenarčič for discussions on related topics and for comments on a preliminary version of the manuscript. We are further grateful to Thomas Wellens, Martin Koppenh\"ofer and Michael Krebsbach for fruitful discussions on the \texttt{ibmq\_ehningen} machine. 
We acknowledge funding from the Deutsche Forschungsgemeinschaft (DFG, German Research Foundation) under Project No. 435696605, through the Research Unit FOR 5413/1, Grant No. 465199066, through the Research Unit FOR 5522/1, Grant No. 499180199, and under Germany’s Excellence Strategy – EXC-Number 2064/1 – Project number 390727645. This project has also received funding from the European Union’s Horizon Europe research and innovation program under Grant Agreement No. 101046968 (BRISQ). F.C.~is indebted to the Baden-W\"urttemberg Stiftung for the financial support of this research project by the Eliteprogramme for Postdocs. This work is funded by the Ministry of Economic Affairs, Labour and Tourism Baden-W\"urttemberg in the frame of the Competence Center Quantum Computing Baden-W\"urttemberg (project ‘QORA II’).

\appendix

\begin{figure*}
    \centering
    \includegraphics{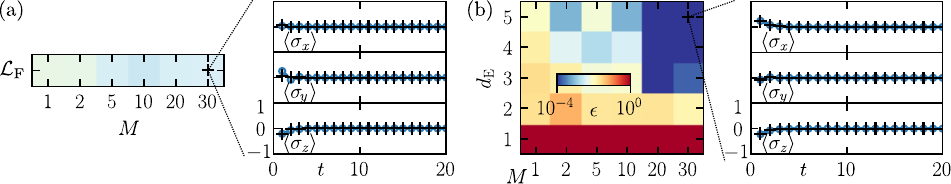}
    \caption{\textbf{Quantum channel without infinitesimal  generator.} (a) Error measure associated with the Lindblad dynamics approximator of Ref.~\cite{Cemin2024} as a function of the number $M$ of considered initial conditions. The plots on the right show a comparison between the predicted expectation values (black crosses) and the exact  simulated data (blue circles). (b) Error measure $\epsilon$ associated with the learned quantum channel as a function of the number $M$ of considered initial conditions and of the environmental dimension $d_{\rm E}$. The panels on the right show a comparison between the predicted expectation values (black crosses) and the exact  simulated data (blue circles).
    }
    \label{fig:fig_s2}
\end{figure*}

\section{Further applications}

\subsection{Quantum channel without infinitesimal generator}
As a further example, we consider the quantum channel
\begin{align}
    \mathcal{E}[\rho] = \frac{\Tr[\rho] \mathds{1} + \rho^T}{3} \, ,
    \label{eq:transpose_like_map}
\end{align}
which implements a single-qubit channel, related to the transposition of the density matrix. 
As discussed in Refs.~\cite{Wolf2008,Wolf2008a}, this quantum channel cannot be described by any (time-dependent) Lindblad master equation.

We follow the same approach as before. We train both the Lindblad dynamics approximator of Ref.~\cite{Cemin2024} [cf.~Fig.~\ref{fig:fig_s2}(a)] and the algorithm presented in the main text [cf.~Fig.~\ref{fig:fig_s2}(b)] for different numbers of considered initial conditions $M$ and, for our new method, for different values of the environmental dimension $d_E$. We then compute the error measure, as defined in Eq.~\eqref{eq:error_function}, using $r=10$ out-of-sample trajectories.
Both the learned quantum channel and the Lindblad dynamics approximator show good agreement with the exact time evolution, as displayed in Fig.~\ref{fig:fig_s2}. Still, the Lindblad approximator cannot accurately capture the whole time evolution, as witnessed by a larger value of the error measure. Conversely, the learned channel is very accurate as soon as $d_{\rm E} \geq 3$, which coincides with the Kraus rank of the exact dynamics \cite{Wolf2008a}, i.e., the minimal number of Kraus operators required for representing the considered channel.

\subsection{Interacting two-qubit system}
In this section, we further establish the performance of our model by learning the dynamics of an interacting open system governed by a time-independent Lindblad master equation. 

\begin{figure*}
    \centering
    \includegraphics{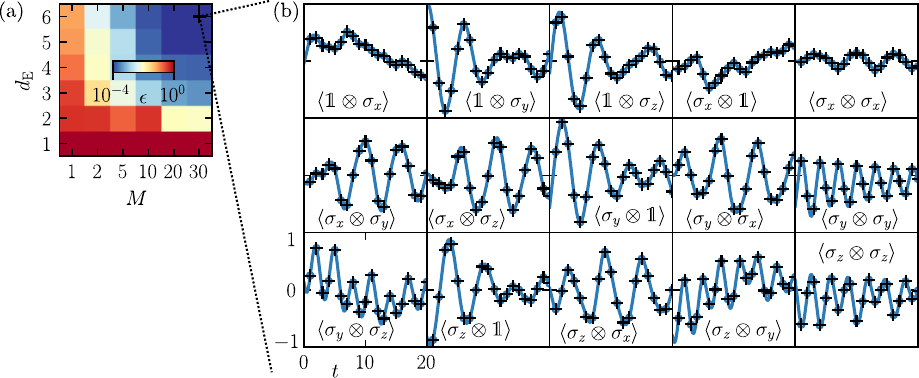}
    \caption{\textbf{Interacting two-qubit system.} We investigate the time-independent Lindblad dynamics \eqref{eq:benchmark} with $V = 0.5 \, \Omega$, $\gamma = 0.01 \, \Omega$ and $\kappa = 0.05 \, \Omega$. We sample and learn the dynamics at stroboscopic times, i.e., multiples of $1 / \Omega$. (a) Error measure as defined in Eq.~\eqref{eq:error_function} in the manuscript for different dimensions of the environment $d_\mathrm{E}$ and different numbers of initial conditions $M$. (b) Predictions of inferred quantum channel (black crosses) for the exact Lindblad dynamics at stroboscopic times (blue dots). }
    \label{fig:fig_s3}
\end{figure*}

Concretely, we consider the dynamics of two driven qubits that interact if both are in the excited state. The open-system dynamics containing local dephasing and decay therefore yields
\begin{align}    
    \label{eq:benchmark}
    \begin{split}
        \dot{\rho} &= \mathcal{L}[\rho]\\
        &= -i[H, \rho] +  \sum_{i = 1}^4 \left\{J_i \rho J_i^\dagger - \frac{1}{2} \left(J_i^\dagger J_i \rho + \rho J_i^\dagger J_i\right) \right\} \, ,
    \end{split}
\end{align}
with Hamiltonian and jump-operators
\begin{align*}
    H &= \frac{\Omega}{2} (\sigma_1^x + \sigma_2^x) + V  n_1 n_2\, , \\
    J_{1,2} &= \sqrt{\gamma} \sigma_{1,2}^- \, , \qquad J_{3,4} = \sqrt{\kappa} \, n_{1,2} \, .
\end{align*}
Here, $\sigma^x_i = \ketbra{0}{1} + \mathrm{h.c.}$, $n_i = \ketbra{1}$, $\sigma^- = \ketbra{0}{1}$ are operators acting on the $i$th qubit. The quantum channel at stroboscipic times, i.e., multiples of $1 / \Omega$, is given by $\mathcal{E} = e^{t \mathcal{L} / \Omega}$.

In Fig.~\ref{fig:fig_s3}, we observe that our method infers the dynamics with arbitrary precision, given sufficiently large $d_{\mathrm{E}}$ and $M$. In consequence, the possibly larger error measure $\epsilon$ in the other examples involving two qubits, originates from non-Markovian effects rather than a possible deficiency of our model to infer dynamics of systems consisting of more than one qubit.

\section{Details on applications in Sec.~\ref{sec:Applications}}
\subsection{\label{sec:reduced_subsystem_dynamics} Reduced subsystem dynamics}
In this section, we extend the discussion related to the reduced subsystem dynamics discussed in Fig.~\ref{fig:subsystem_dynamics}. We further investigate the behavior of the algorithm for the two parameter pairs $(\varphi_x, \varphi_{nn}) = (0.1, 0.1)$ and $(\varphi_x, \varphi_{nn}) = (0.5, 1)$ (already shown in the main text), together with the pair $(\varphi_x, \varphi_{nn}) = (0.5, 2)$. For all these cases, we choose the initial conditions as a product state of Haar random states for the subsystem qubits and of states $\ket{0}$ for all other qubits.

\begin{figure*}[t]
    \centering
    \includegraphics{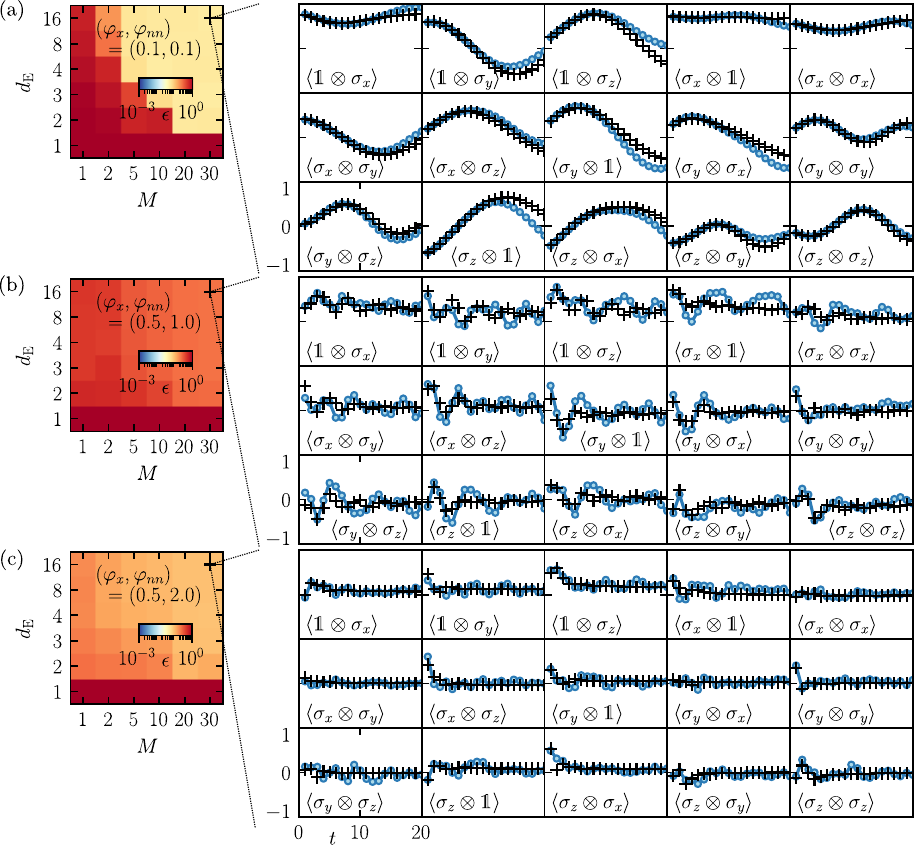}
    \caption{\textbf{Reduced subsystem dynamics.} We report the error measure [Eq.~\eqref{eq:error_function}] and all 15 non-trivial observables for the parameter pairs, (a): $(\varphi_x, \varphi_{nn}) = (0.1, 0.1)$, (b): $(\varphi_x, \varphi_{nn}) = (0.5, 1)$ and (c): $(\varphi_x, \varphi_{nn}) = (0.5, 2)$. Each model is trained on trajectories from $M=30$ different initial conditions and the error measure is computed using $r=10$ out-of-sample trajectories, one of which is reported on the right in each panel.
    The black crosses represent the prediction of our model, while blue circles correspond to the simulated data.}
    \label{fig:fig_s4}
\end{figure*}

Let us first discuss the two parameter pairs in Fig.~\ref{fig:fig_s4}, also investigated in the main text. For $(\varphi_x, \varphi_{nn}) = (0.1, 0.1)$ [panel (a)], the predicted dynamics accurately reproduces the data up to $t \approx 10$, before deviating both quantitatively and qualitatively (see, e.g., $\expval{\sigma_y \otimes \sigma_x}$). In contrast to this, the dynamics for $(\varphi_x, \varphi_{nn}) = (0.5, 1)$ [panel (b)] shows notable differences already at early times. For the additional parameters $(\varphi_x, \varphi_{nn}) = (0.5, 2)$ [panel (c)], we observe a good agreement between our prediction and the simulated data. This behavior may be attributed to a faster scrambling in the circuit which quickly drives the system toward the fully mixed state. The latter can be accurately mimicked by the learned quantum channels. However, small deviations from this mixed state are not fully captured by the learned quantum channel. 

\subsection{\label{sec:ibmq_ehningen} Dynamics on the IBM quantum processor}
In this section, we discuss the data obtained from the \texttt{ibmq\_ehningen} quantum processor (accessed on the 3rd of January 2024), and make connections with the discussion related to Fig.~\ref{fig:ibmq_ehningen} in the main text. The initial conditions $\rho(0) = \ketbra{\psi(0)}{\psi(0)}$ are such that all qubits, but the two composing the subsystem of interest, are initialized in state $\ket{0}$. The initial states for the two-qubit subsystem are taken in product form and such that half of the times they are  formed by a same random state for the two qubits, while the other half of the times the two qubits are characterized by two different random states. Single-qubit random states are sampled from the Haar measure \cite{Meckes2019,Qiskit2019}. 

\begin{figure*}
    \centering
    \includegraphics{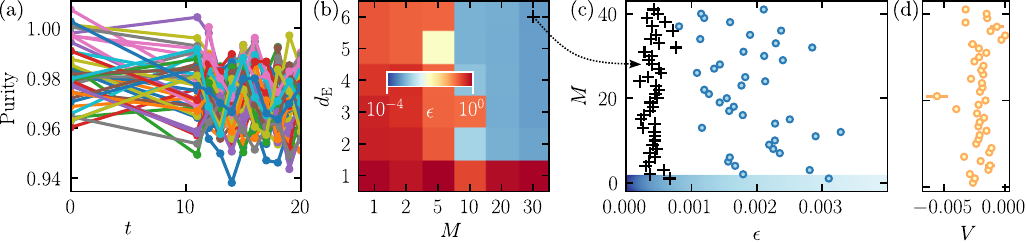}
    \caption{\textbf{Parallel $\sqrt{X}$ dynamics on \texttt{ibmq\_ehningen} - purity and  error measure.} (a) Purity $\Tr{(\rho(t))^2}$ for all the different trajectories (here represented with different colors) at all measured times on the \texttt{ibmq\_ehningen}. (b) Error measure for the learned channel as a function of $M$ and $d_{\rm E}$. (c) Using the best learned quantum channel [top right corner in panel (b)], we compute the error measure $\epsilon$ against $M$ different single trajectories, shown through the black crosses. We also compute the error measure in Eq.~(\ref{eq:error_function}) associated with the ideal $\sqrt{X}$-gate (blue circles). For reference, in panel (c) we report the color scale corresponding to the one in Fig.~\ref{fig:ibmq_ehningen}. (d) For each of the trajectories in panel (c), we report the parameter $V$ characterizing the $ZZ$-coupling obtained by fitting the Pearson correlation coefficients with the ansatz in Eq.\eqref{eq:ZZ_coupling_Hamiltonian}.}
    \label{fig:fig_s5}
\end{figure*}
Let us first consider the purity presented in Fig.~\ref{fig:fig_s5}(a). For unitary dynamics, perfect readout, and infinitely many shots, the purity would be equal to 1. Here, as we are dealing with dynamics on an actual quantum device, we observe a decrease of the purity, although the latter remains  close to 1 throughout. 
In Fig.~\ref{fig:fig_s5}(b), we report the error measure as already shown in Fig.~\ref{fig:ibmq_ehningen}. Out of all the learned models, we pick the best (top right corner), and compute the error measure $\epsilon$ for each simulated initial condition $M$, shown in panel (c) as black crosses. For comparison, we compute the error measure using the ideal $\sqrt{X}$ dynamics as quantum channel (blue circles). Despite the error for the ideal $\sqrt{X}$ gate being quite small for all the considered initial conditions, our algorithm can better approximate the actual dynamics.
Let us mention, that the model was trained using the first 30 trajectories in panel (c), see the $y$-axis. Despite this, the error measure does not show an increase for the last 10 trajectories, indicating a good extrapolation capability of the model.
In Fig.~\ref{fig:fig_s5}(d), we provide the $ZZ$-coupling strength $V$ that we extract by fitting the Pearson correlation coefficients using the Hamiltonian in Eq.~\eqref{eq:ZZ_coupling_Hamiltonian}. Based on this observation, we choose $V \in (-0.003, -0.001)$ for the comparison in the main text.

\begin{figure*}
    \centering
    \includegraphics{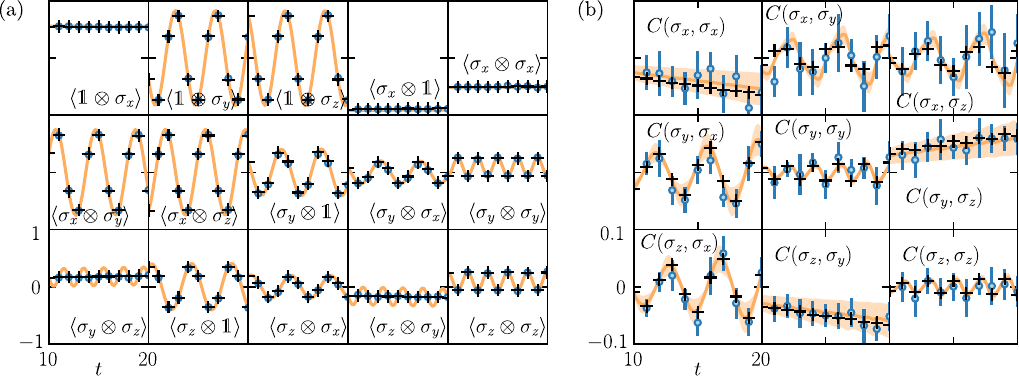}
    \caption{\textbf{Parallel $\sqrt{X}$ dynamics on \texttt{ibmq\_ehningen} - predicted trajectories and Pearson correlation coefficient.} (a) For the best learned quantum channel in Fig.~\ref{fig:ibmq_ehningen}, top right corner, we compare its prediction (black crosses) with the measured expectation values (blue circles). (b) We provide all Pearson correlation coefficients associated with the time-evolution in (a). In both panels, we also provide the heuristic prediction of Eq.~\eqref{eq:ZZ_coupling_Hamiltonian} with $V \in (-0.003, -0.001)$ (yellow shaded region).}
    \label{fig:fig_s6}
\end{figure*}

In Fig.~\ref{fig:fig_s6}, we provide the full comparison between the prediction of our model against the measurements on the \texttt{ibmq\_ehningen} device. The prediction (black crosses) agree with dynamics on the device (blue circles) within statistical errors both for the expectation values [see panel (a)] and for the Pearson correlation coefficients [see panel (b)]. We note that the $\sqrt{X}$ gate is already nearly perfect rendering an accurate prediction of the Pearson correlation coefficient a challenging task for any noise model. 

\FloatBarrier

\bibliography{biblio.bib}

\end{document}